# Cosmology with One Galaxy: Autoencoding the Galaxy Properties Manifold

Amanda Lue[1], Shy Genel[2,3], Marc Huertas-Company[4,5,6], Francisco Villaescusa-Navarro[2,7], and Matthew Ho[1,8]
[1] Department of Astronomy, Columbia University, New York, NY 10027, USA
[2] Center for Computational Astrophysics, Flatiron Institute, 162 Fifth Avenue, New York, NY 10010, USA
[3] Columbia Astrophysics Laboratory, Columbia University, 550 West 120th Street, New York, NY 10027, USA
[4] Instituto de Astrofísica de Canarias, c/Vía Láctea sn, 38025 La Laguna, Spain
[5] Universidad de La Laguna. Avda. Astrofísico Fco. Sanchez, La Laguna, Tenerife, Spain
[6] LERMA, Observatoire de Paris, CNRS, PSL, Université Paris-Cité, Paris, France
[7] Department of Astrophysical Sciences, Princeton University, Peyton Hall, Princeton, NJ 08544, USA
[8] Institut d'Astrophysique de Paris, CNRS, Sorbonne Université, France

## Abstract

Cosmological simulations like CAMELS and IllustrisTNG characterize hundreds of thousands of galaxies using various internal properties. Previous studies have demonstrated that machine learning can be used to infer the cosmological parameter $\Omega_m$ from the internal properties of even a single randomly selected simulated galaxy. This ability was hypothesized to originate from galaxies occupying a low-dimensional manifold within a higher-dimensional galaxy property space, which shifts with variations in $\Omega_m$. In this work, we investigate how galaxies occupy the high-dimensional galaxy property space, particularly the effect of $\Omega_m$ and other cosmological and astrophysical parameters on the putative manifold. We achieve this by using an autoencoder with an information-ordered bottleneck, a neural layer with adaptive compression, to perform dimensionality reduction on individual galaxy properties from CAMELS simulations, which are run with various combinations of cosmological and astrophysical parameters. We find that for an autoencoder trained on the fiducial set of parameters, the reconstruction error increases significantly when the test set deviates from fiducial values of $\Omega_m$ and $A_{\mathrm{SN1}}$, indicating that these parameters shift galaxies off the fiducial manifold. In contrast, variations in other parameters such as $\sigma_8$ cause negligible error changes, suggesting galaxies shift along the manifold. These findings provide direct evidence that the ability to infer $\Omega_m$ from individual galaxies is tied to the way $\Omega_m$ shifts the manifold. Physically, this implies that parameters like $\sigma_8$ produce galaxy property changes resembling natural scatter, while parameters like $\Omega_m$ and $A_{\mathrm{SN1}}$ create unsampled properties, extending beyond the natural scatter in the fiducial model.

## 1. Introduction

The CAMELS project (F. Villaescusa-Navarro et al. 2021) investigates the link between cosmological parameters and astrophysical observations through a suite of cosmological simulations containing hundreds of thousands of galaxies, each characterized by various internal properties. A key goal of this project is to determine if observations of small-scale structures, such as galaxies, can be used to infer cosmological parameters. To that end, F. Villaescusa-Navarro et al. (2022) trained a neural network on galaxy properties from the CAMELS simulations to perform likelihood-free inference on $\Omega_m$, $\sigma_8$, and four astrophysical parameters. They found that using the internal properties of even a single galaxy, $\Omega_m$ could be constrained to about 10% accuracy when $\Omega_b$ is fixed.

Further analysis showed that $\Omega_m$ is not directly linked to any one galaxy property. Instead, F. Villaescusa-Navarro et al. (2022) hypothesized that correlations between galaxy properties and $\Omega_m$ suggested the existence of a low-dimensional manifold within the higher-dimensional galaxy properties space, which shifts in response to changes in $\Omega_m$. Consequently, the position of a single randomly selected galaxy within this galaxy properties space is sufficient for identifying the $\Omega_m$-dependent manifold, thus allowing inference of the $\Omega_m$ value.

This manifold organizes galaxy properties in a structured and meaningful way, where galaxies close to each other share similar characteristics. While changes in one property can be related to others in complex, high-dimensional ways, this intricate organization underlies simpler, well-known 2D scaling relations like the Tully–Fisher relation (TFR), which connects circular velocity and luminosity in galaxies (R. B. Tully & J. R. Fisher 1977; V. Avila-Reese et al. 2008; M. Cappellari 2015; S. M. Fall & A. J. Romanowsky 2018). There is a rich history of studying galaxy scaling relations, which continues to evolve today (A. A. Dutton & F. C. van den Bosch 2009; J. Fu et al. 2010; Y.-j. Peng & R. Maiolino 2014). Recognizing that these relations may be projections of a thin, low-dimensional manifold that shifts with cosmological parameters in a high-dimensional space offers a compelling new perspective.

Although these insights are derived from simulations, the concept of galaxy properties residing on a manifold can extend to real observations. Characterizing this manifold could reduce uncertainties in observed galaxy properties by enforcing a manifold structure, potentially improving the accuracy of measurements in the real Universe (F. Villaescusa-Navarro et al. 2022; S. Cooray et al. 2023). Furthermore, a data-driven characterization of this manifold would provide a valuable reference for analytical models of galaxy formation, advancing our understanding of the field (R. S. Somerville & J. R. Primack 1999).

In this study, we further support the manifold interpretation proposed by F. Villaescusa-Navarro et al. (2022) by investigating how variations in six parameters from the original CAMELS-IllustrisTNG simulations, $\Omega_m$, $\sigma_8$, $A_{\rm SN1}$, $A_{\rm AGN1}$, $A_{\rm SN2}$, and $A_{\rm AGN2}$, influence the way galaxies occupy the high-dimensional galaxy properties space. To explore this, we employ an autoencoder neural network equipped with an information-ordered bottleneck (IOB) layer, which enforces structured, hierarchical encoding of latent variables (M. Ho et al. 2023). During training, the network minimizes the reconstruction loss, measuring how well the compressed representation captures the critical features of the data. The IOB layer prioritizes encoding the most informative features in the top latent variables, optimizing the network to capture information essential for reconstructing input data in these top variables. As the network compresses the galaxy properties into a latent space, it learns a lower-dimensional, nonlinear representation of the data. While this latent space relates to the underlying galaxy properties manifold, its structure is complex and not easily deciphered (A. McGovern et al. 2019; L. Maglianella et al. 2023). However, our ability to reconstruct the galaxy properties from this latent space provides direct evidence for the existence of the hypothesized galaxy manifold.

Furthermore, to examine how variations in $\Omega_m$, as well as other cosmological and astrophysical parameters, influence the galaxy manifold, we use galaxy properties from three sets within the IllustrisTNG (TNG; D. Nelson et al. 2019) suite of the CAMELS simulations (F. Villaescusa-Navarro et al. 2021). The CAMELS Cosmic Variance (CV) set, where all parameters are fixed at fiducial values, serves as a baseline for comparison. The CAMELS Latin Hypercube (LH) set, which simultaneously varies cosmological and astrophysical parameters, allows us to explore the effects of multiple parameter changes. Lastly, the CAMELS 1-parameter-at-a-time (1P) set isolates the impact of individual parameters by varying them one at a time while holding others constant. Building on the manifold interpretation of the results from F. Villaescusa-Navarro et al. (2022), we analyze changes in the root mean squared error (RMSE) of reconstruction to investigate how these parameter variations might influence the manifold. Our results reveal that deviations in $\Omega_m$ and $A_{\rm SN1}$ strongly correlate with increased RMSE, suggesting that these parameters shift galaxies off the fiducial manifold. In contrast, changes in the other parameters, $\sigma_8$, $A_{\rm AGN1}$, $A_{\rm SN2}$, and $A_{\rm AGN2}$, result in negligible RMSE variations, indicating that shifts in galaxy properties occur along the manifold rather than off of it. These findings provide direct evidence for the manifold hypothesis and explain why $\Omega_m$ can be inferred from a single galaxy's properties, linking this ability to how $\Omega_m$ shifts the manifold itself.

We also provide a physical interpretation for the manifold shift: variations in $\Omega_m$ and $A_{\rm SN1}$ generate galaxy properties that are unsampled in the fiducial training set, extending the range of galaxy characteristics beyond the natural scatter. In contrast, changes in parameters such as $\sigma_8$, lead to variations in galaxy properties that mirror the natural scatter already represented in the training set. This distinction underscores the unique influence of $\Omega_m$ and $A_{\rm SN1}$ on galaxy characteristics and their role in shaping the underlying manifold.

This paper is structured as follows. Section 2 describes the data and outlines the network architecture. In Section 3, we present our findings, followed by a discussion of their physical interpretation in Section 4. Finally, Section 5 summarizes the key conclusions and proposes directions for future research.

## 2. Methods

### 2.1. Simulations

The models were trained on the galaxy properties of individual galaxies from the IllustrisTNG (TNG; D. Nelson et al. 2019) suite of the CAMELS simulations (F. Villaescusa-Navarro et al. 2021). While IllustrisTNG represents a cutting-edge suite of large-volume cosmological simulations focused on galaxy formation, CAMELS consists of numerous smaller-volume simulations designed as perturbations on the cosmological parameters from the fiducial IllustrisTNG simulations.

This work focuses on the CAMELS-IllustrisTNG suite (F. Villaescusa-Navarro et al. 2021), which is run using the same AREPO code and subgrid physics used in IllustrisTNG simulations (V. Springel 2010). CAMELS includes other simulation suites, such as SIMBA (R. Davé et al. 2019), and although we do not analyze them directly in this work, we expect that our conclusions would generalize to them. This expectation follows from the findings of F. Villaescusa-Navarro et al. (2022), which demonstrated that $\Omega_m$ could be inferred from galaxies in both the IllustrisTNG and SIMBA simulation suites.

The CAMELS box size is 25 Mpc, and each simulation has a comoving volume of $(25\,{\rm Mpc}/h)^3$, containing $256^3$ dark matter particles. The dark matter particles have a mass of $6.49 \times 10^7(\Omega_m - \Omega_b)/0.251 h^{-1} M_\odot$. The simulations are hydrodynamic and include an additional $256^3$ gas particles, which have an initial mass of $1.27 \times 10^7 \Omega_b/0.049 h^{-1} M_\odot$. Within the IllustrisTNG suite, the simulations are divided into different sets. The sets used in this study share the same resolution and have $\Omega_b = 0.049$. The three main sets employed are:

1. The LH set, which comprises 1000 simulations characterized by concurrent variations in cosmological and astrophysical parameters, arranged in an LH.
2. The CV set, which consists of 27 simulations that share the same cosmological and astrophysical parameters but differ in the random seeds used for their initial conditions.
3. The 1P set, which includes 61 simulations where cosmological and astrophysical parameter variations occur within the same range as the LH set. However, in the 1P set, only one parameter changes at a time while keeping all other parameters at their fiducial value.

Furthermore, the parameter ranges in these simulations implicitly determine our priors. These ranges explore a physically motivated space of parameter variations, ensuring that the sampled values reflect plausible astrophysical and cosmological conditions. This ensures a fair basis for comparing parameter effects within the studied range.

### 2.2. Data Preprocessing

For our data set, we independently prepare each of the three CAMELS simulation sets. We include only subhalos with a stellar mass greater than $10^8 M_\odot$. Subhalo identification is achieved using the friends-of-friends (J. P. Huchra & M. J. Geller 1982; M. Davis et al. 1985) and SUBFIND algorithms (V. Springel et al. 2001). We focus on the galaxy properties used in F. Villaescusa-Navarro et al. (2022), with descriptions outlined in Table 1. Notably, we exclude the galaxy's peculiar velocity from the input properties, as its minimal contribution to the mean squared error (MSE) indicated it was not essential for characterizing the manifold. Removing this property led to a reduction in the overall MSE without omitting any critical information.

**Table 1**
Galaxy Properties and Descriptions

| Galaxy Property | Description |
|---|---|
| $M_{\rm BH}$ | Black hole mass of the galaxy |
| $M_{\rm t}$ | Total mass of the galaxy's subhalo, including dark matter, gas, stars, and black holes |
| $R_{\rm t}$ | Radius containing half of the total mass of the galaxy's subhalo |
| $R_*$ | Radius containing half of the galaxy's stellar mass |
| $M_{\rm g}$ | Total gas mass of the galaxy, including the circumgalactic medium |
| $M_*$ | Stellar mass of the galaxy |
| $V_{\rm max}$ | Maximum circular velocity of the galaxy's subhalo, calculated as $V_{\rm max} = \max(\sqrt{GM(<R)/R})$ |
| $\sigma_v$ | Velocity dispersion of all particles within the galaxy's subhalo |
| $R_{\rm max}$ | Radius at which $\sqrt{GM(<R_{\rm max})/R_{\rm max}} = V_{\rm max}$ |
| $Z_{\rm g}$ | Mass-weighted gas metallicity of the galaxy |
| $Z_*$ | Mass-weighted stellar metallicity of the galaxy |
| SFR | Star formation rate of the galaxy |
| $U$ | Galaxy magnitude in the $U$ band |
| $K$ | Galaxy magnitude in the $K$ band |
| $g$ | Galaxy magnitude in the $g$ band |
| $J$ | Magnitude of the galaxy's subhalo spin vector |

After applying the stellar mass cut, we impute zero-valued data points by adding a small positive constant to prevent NaN errors. The data, excluding color magnitude properties (already expressed as logarithmic luminosity values), were then transformed using a base-10 logarithm to mitigate the effects of differing value ranges across properties. Next, we applied standard scaling (F. Pedregosa et al. 2011) using the mean and standard deviation calculated from the LH set, which exhibits the greatest variance. The processed data set was subsequently divided into training, validation, and test sets, comprising 60%, 20%, and 20% of the total data, respectively. To prevent information leakage, simulations were randomly shuffled, and galaxies from the same simulation were confined to a single set. It is important to note that, for the CAMELS LH set, we utilized only the first 250 simulations instead of the full 1000 due to computational limitations, resulting in a training set of approximately 175,500 galaxies for the LH model and 14,000 for the CV model. However, since the simulations are arranged in an LH and are not ordered in any particular way, this selection effectively represents a random subset. As a result, it does not systematically truncate the prior range or introduce bias in the parameter coverage.

### *2.3. Architecture*

A key component in our model is the IOB layer developed by M. Ho et al. (2023). In the IOB layer, data encoding occurs hierarchically, prioritizing the most significant information in the top latent variables. During training, the bottleneck size is dynamically adjusted by masking the outputs of the nodes. The nodes are opened in a fixed order, ensuring that if the third node is open, then the first and second must also be open. Consequently, the top nodes contribute the most to the log-likelihood as they are always open and contributing to the loss function. This structure incentivizes the network to pass priority information through the top nodes, producing ordered latent variables.

Our network takes as its input the 16 individual galaxy properties. The goal is to train the network to reconstruct input data by minimizing the MSE loss, aiming for high fidelity in data reconstruction. In training the model we are computing the following (M. Ho et al. 2023):

$$\hat{\theta} = \arg\max_{\theta\in\Theta} \sum_{i=1}^{N}\sum_{k=0}^{k_{\rm max}} \rho_k \ell[f_\theta^{(k)}(x_i), x_i]. \quad (1)$$

Here, $\hat{\theta}$ represents the set of parameters that allow the network to best reconstruct the input data from its latent representation. The capital $\Theta$ refers to the set of all possible values for the parameter vector $\theta$, which defines the model's parameters. The input $x_i$ is related to the output through the function $f_\theta^{(k)}(x_i)$. The log-likelihood of the model, denoted as $\ell$, is computed for various bottleneck widths $k$. This likelihood can be interpreted as an implicit likelihood that is learned during training, as the neural network learns to encode the data structure through its latent representation. However, the likelihood expression used here is defined in terms of reconstruction accuracy. Specifically, the log-likelihood $\ell$ of our autoencoder outputs is defined as a Gaussian distribution, with fixed variance and mean of the true input $x_i$. Thus, $\ell \propto \|f_\theta^{(k)}(x_i) - x_i\|_2^2$, also known as the vector MSE between the truth and reconstruction. Each log-likelihood term is weighted by a constant scalar $\rho_k$, and these weighted log-likelihoods are summed over $N$ individual data points, considering all $k$ from 0 to $k_{\rm max}$ (inclusive). The maximum bottleneck, $k_{\rm max} = 16$, matches the number of input properties, effectively capturing information across all bottleneck widths. This formulation effectively condenses and encodes essential information efficiently.

Hyperparameters, including the number of fully connected layers, learning rate, neuron count per layer, and optimizer, were selected through experimentation to optimize loss convergence. However, minimal tuning was required, as deviations from the default hyperparameters resulted in only marginal differences in loss convergence. Our network follows a forward-pass structure with a defined encoder–decoder configuration, where each component consists of four fully connected layers with ReLU activations. Loss is computed using MSE, with a batch size of 32. We define convergence based on a minimum change of 0.0001 in validation loss sustained for at least 20 epochs.

## 3. Results: The Impact of Parameter Variation on Reconstruction

We begin by examining whether a model trained on the CV set, where cosmological and astrophysical parameters are fixed at fiducial values, can successfully reconstruct galaxies from the LH set, in which parameters vary simultaneously, and vice versa. Two models using the same architecture and hyperparameters are trained for this purpose: one on galaxy properties from the CV set and another on the LH set. Figure 1 is a representation of how well the network reconstructs the test set with $k$ bottlenecks open. The MSE is normalized by the population variance, which quantifies the variability across galaxy properties, thereby improving comparability. The model trained on the CV set performs worse when tested on galaxies from the LH set, displaying higher reconstruction errors. On the other hand, the LH-trained model shows similar reconstruction errors for both test sets, with performance comparable to the CV-trained model when tested on CV galaxies. From the results we present later, it becomes evident that the elevated reconstruction error of the CV model on LH galaxies is a consequence of $\Omega_m$ shifting galaxies off the fiducial manifold learned by the model.

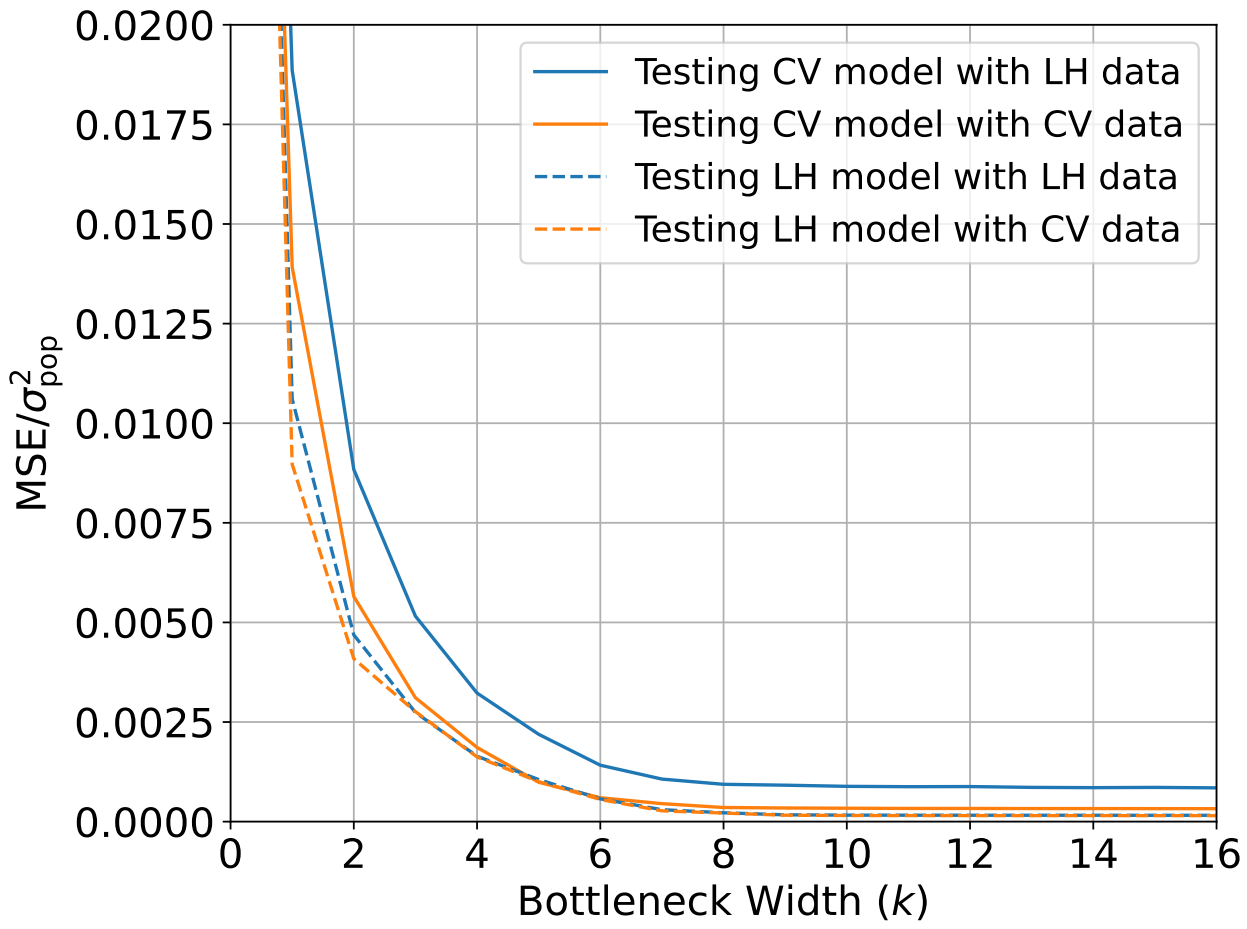

**Figure 1.** MSE of the reconstruction for each bottleneck ($k$) width normalized by the population variance. The color denotes the test data, and the line style denotes the data used to train the model. The model trained on CV data (solid) has a higher MSE when tested on data from the LH set, while the LH model (dotted) performs similarly when tested on either data set.

To further investigate this behavior, we examine the reconstruction error of the CV model at a fixed bottleneck, $k = 5$, when tested on galaxies from the 1P simulation sets. Figure 2 shows the relationship between RMSE and each of the six parameters varied in the 1P simulations. A model trained on galaxies from the CV simulations learns to reconstruct galaxies produced by a specific cosmology with the fiducial values of all six parameters. RMSE increases significantly with greater deviation from the values of $\Omega_m$ and $A_{\rm SN1}$ used in training. In contrast, changes to $\sigma_8$, $A_{\rm AGN1}$, $A_{\rm SN2}$, and $A_{\rm AGN2}$ do not lead to noticeable changes in RMSE. These findings suggest that only $\Omega_m$ and $A_{\rm SN1}$ shift galaxies off the fiducial galaxy properties manifold. This effect is visually demonstrated in Figure 3, which illustrates how variations in $\Omega_m$ and $\sigma_8$ alter the structure of the galaxy property manifold. As these parameters deviate from their fiducial values, galaxies are pushed further from the manifold, making it harder for the network trained on the fiducial set to reconstruct them, resulting in higher RMSE. While changes in the other cosmological and astrophysical parameters are known to affect galaxy properties (R. J. McGibbon & S. Khochfar 2023; V. Busillo et al. 2025), their lack of correlation with RMSE suggests that these effects occur within the fiducial manifold, rather than shifting galaxies away from it.

The influence of $\Omega_m$ aligns with the results reported by F. Villaescusa-Navarro et al. (2022), which demonstrate the constraining power of their models on $\Omega_m$ using a single galaxy. While F. Villaescusa-Navarro et al. (2022) note little constraining power on $A_{\rm SN1}$, the work of C. Chawak et al. (2024) indicates that $A_{\rm SN1}$ can be constrained when multiple galaxies are used. This observation is consistent with our findings, where the parameters constrained by these models are the same as those that correlate with RMSE in our study.

Even though both $\Omega_m$ and $A_{\rm SN1}$ shift the learned manifold, $\Omega_m$ induces a distinct transformation in the 16-dimensional property space that allows the network to distinguish its effects from those of $A_{\rm SN1}$. This interpretation explains why a single galaxy can still be used to infer $\Omega_m$, as demonstrated by F. Villaescusa-Navarro et al. (2022).

Figure 4 illustrates a similar trend, showing the results of testing on the LH set depicted in Figure 1. The relationship between $\Omega_m$ and RMSE in Figure 4 mirrors the general pattern observed in Figure 2, where only $\Omega_m$ is varied. Since multiple parameters are varied in this case, examining RMSE as a function of $\Omega_m$ naturally introduces a noticeable scatter. In particular, the color gradient highlights the influence of $A_{\rm SN1}$, which is known to affect RMSE and therefore contributes to the observed scatter.

To assess the impact of outliers, we calculate both the mean and median reconstruction errors over all 1P galaxies, as shown in Figure 2. We also examined the distribution of reconstruction errors for individual galaxies from each simulation (details not shown). The minor differences between the mean and median reflect a slight asymmetry in the distribution, which deviates from a strict Gaussian shape. Nevertheless, the mean is not significantly influenced by outliers such as the most massive galaxies. N. Echeverri et al. (2023) investigate the impact of outliers on the prediction of $\Omega_m$ and find that removing outliers markedly improves model accuracy. However, in our case, outliers appear to have minimal effect, as evidenced by the close alignment of the mean and median.

## 4. Discussion

### 4.1. Physical Interpretation of Manifold Shifts

To understand why certain parameters affect the reconstruction error, we first consider the physical changes each parameter induces (F. Villaescusa-Navarro et al. 2021). $\sigma_8$ is associated with the clustering of matter, and altering it impacts halo formation times. However, the fiducial model (CV set) already samples a range of halo formation times. As a result, when the network trained on the CV set encounters galaxies from simulations with different values of $\sigma_8$, it has already seen similar galaxies in the fiducial set and can reconstruct them with similar errors. In contrast, $\Omega_m$ directly affects the amount of dark matter and gas in each halo, producing galaxies that are fundamentally different from those in the CV set. The key difference is that there is no analogous effect of changing $\Omega_m$ within the fiducial simulations, meaning the network has not been exposed to galaxies like those created under different $\Omega_m$ conditions.

To further illustrate the impact of $\Omega_m$, we can examine the TFR as a lower-dimensional projection of the galaxy properties manifold. Intuitively, altering $\Omega_m$ would shift the TFR, which connects circular velocity and luminosity or stellar mass. For a 3D visualization of the TFR and how it shifts with varying $\Omega_m$, see Figure 7 of F. Villaescusa-Navarro et al. (2022). Galaxies span a range of masses regardless of the value of $\Omega_m$, but in simulations with higher $\Omega_m$, we expect a given stellar mass to reside in a more massive halo, implying a deeper potential well. However, merely a change in the TFR does not necessarily imply that the higher-dimensional manifold itself has changed. In principle, galaxies could potentially slide along the manifold in some third dimension. Yet, as we noted earlier, there is no natural scatter in the fiducial model that behaves like $\Omega_m$ variations, suggesting that $\Omega_m$ truly shifts the manifold rather than galaxies moving within it.

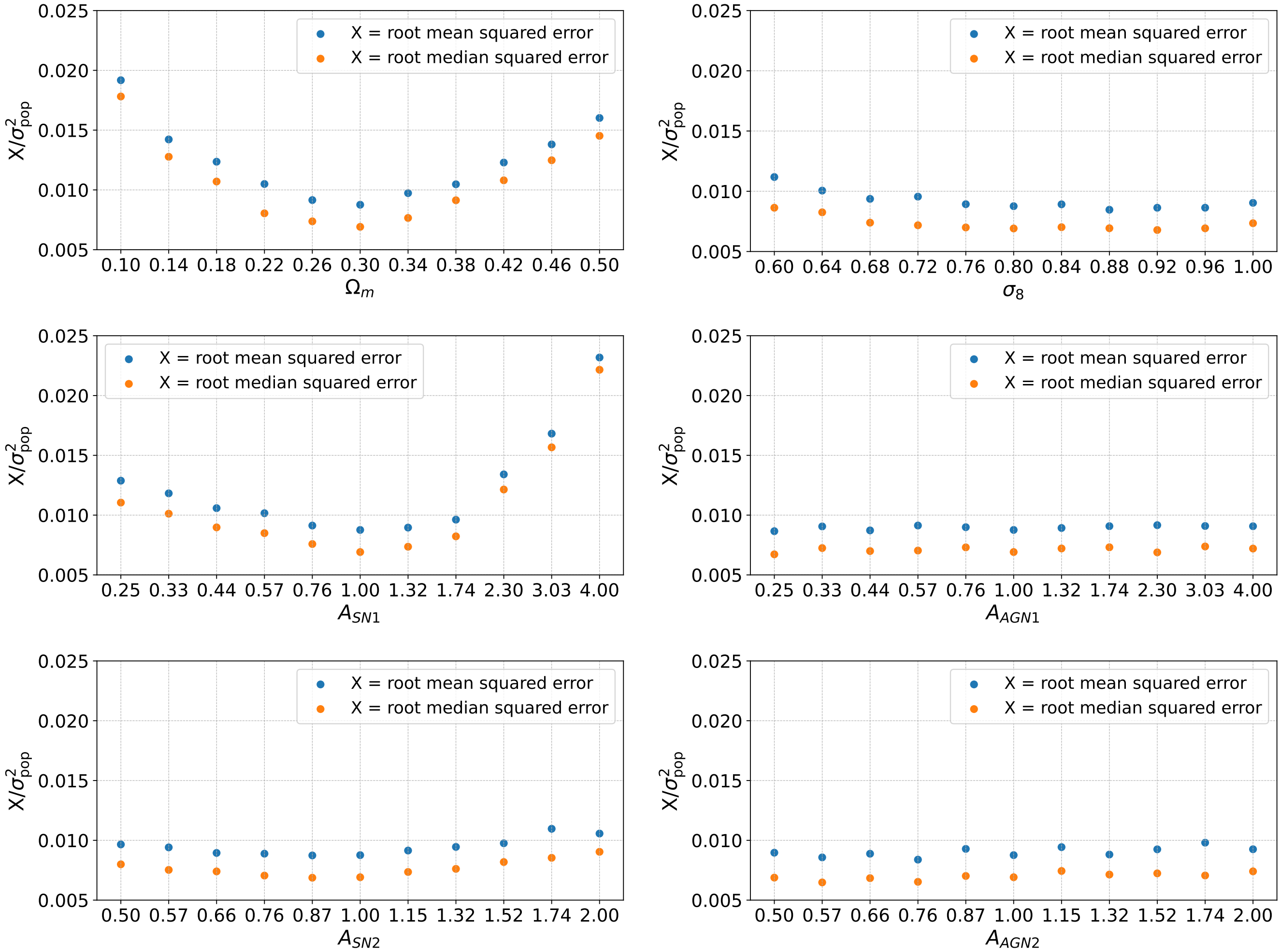

**Figure 2.** Impact of six cosmological and astrophysical parameters on normalized RMSE (blue points) and median (orange points) squared error. Errors are calculated across six sets of 11 CAMELS 1P simulations, each varying one parameter: $\Omega_m$, $\sigma_8$, $A_{\rm SN1}$, $A_{\rm AGN1}$, $A_{\rm SN2}$, and $A_{\rm AGN2}$. The model, trained on the CAMELS CV set, shows RMSE dependence on $\Omega_m$ and $A_{\rm SN1}$, while the other parameters have negligible effect.

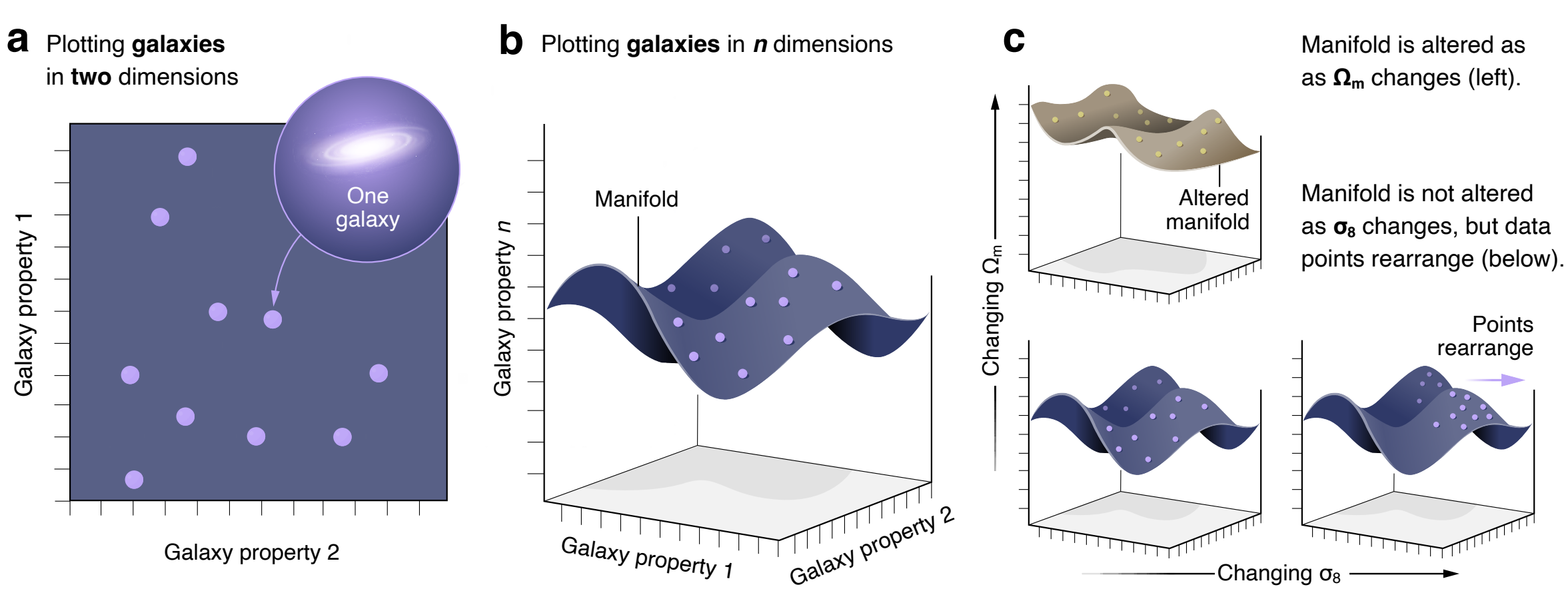

**Figure 3.** Visualization of the galaxy property manifold and the effect of changing $\sigma_8$ and $\Omega_m$.

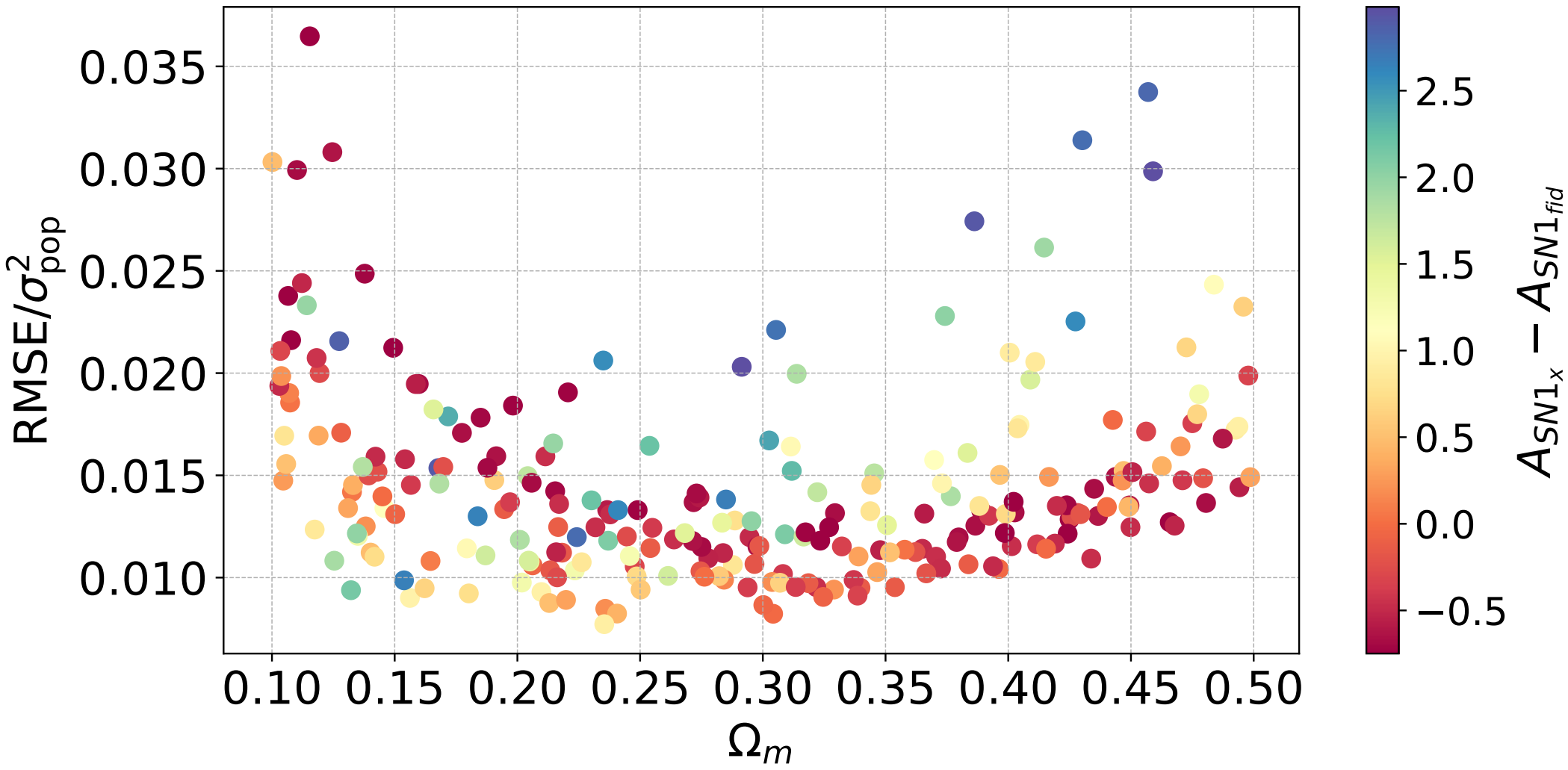

**Figure 4.** The impact of $\Omega_m$ on the normalized RMSE. Errors are calculated across 250 CAMELS LH simulations, where six parameters, $\Omega_m$, $\sigma_8$, $A_{\rm SN1}$, $A_{\rm AGN1}$, $A_{\rm SN2}$, and $A_{\rm AGN2}$, are varied simultaneously. The model, trained on the CAMELS CV set, exhibits an RMSE pattern consistent with that shown in the first panel of Figure 2. The color bar represents the deviation of the $A_{\rm SN1}$ value in each simulation from its fiducial value of 1. These deviations are correlated with the RMSE and contribute to the observed scatter.

We can apply similar reasoning to understand the influence of astrophysical parameters, although our physical interpretation is less developed. $A_{\rm SN1}$ regulates the energy of galactic winds per unit star formation rate. Modifying $A_{\rm SN1}$ leads to galaxies with outflows that differ significantly from those in the fiducial set, causing the network to struggle in reconstruction. In contrast, $A_{\rm SN2}$ affects galactic wind speed, but this strongly depends on local velocity dispersion, a property that already varies naturally in the fiducial model. As a result, changes in $A_{\rm SN2}$ likely shift galaxies along the fiducial manifold, which could explain the lack of correlation between $A_{\rm SN2}$ and the reconstruction error.

### 4.2. Latent Variables

The IOB layer encodes the most critical information in the top latent variables. To explore what these variables might represent, we compute the correlation matrix between the 16 latent variables and the 16 individual galaxy properties. This analysis employs the Pearson correlation coefficient, which quantifies how well the relationship between these two sets of data can be described by a linear equation (D. Freedman et al. 2007). This method highlights linear correlations, but weak linear correlations do not necessarily mean a latent variable is unrelated to physical properties. Therefore, it is important to acknowledge that nonlinear associations may exist and are not captured by this covariance analysis. Nevertheless, for simplicity, we focus solely on linear correlations.

Figure 5 demonstrates that the top latent variable correlates strongly with stellar mass and other mass-related properties, such as metallicity. For instance, black hole mass, often tied to both total and stellar mass, also follows this trend, with galaxies of higher stellar and total masses hosting more massive black holes. Supermassive black holes at galaxy centers further influence dynamics and star formation rates (P. Podigachoski et al. 2015). Since stellar mass reflects a galaxy's star formation history and the available gas for star formation, it is unsurprising that the top latent variable strongly correlates with various mass-related properties. Stellar mass and star formation rate are closely linked (A. R. Tomczak et al. 2016), driven by factors such as gas content and interstellar medium density, which are also reflected in stellar and gas metallicity. Furthermore, the $K$, $g$, and $U$ magnitudes show strong negative correlations with the top latent variable, as more massive stars exhibit smaller apparent magnitudes.

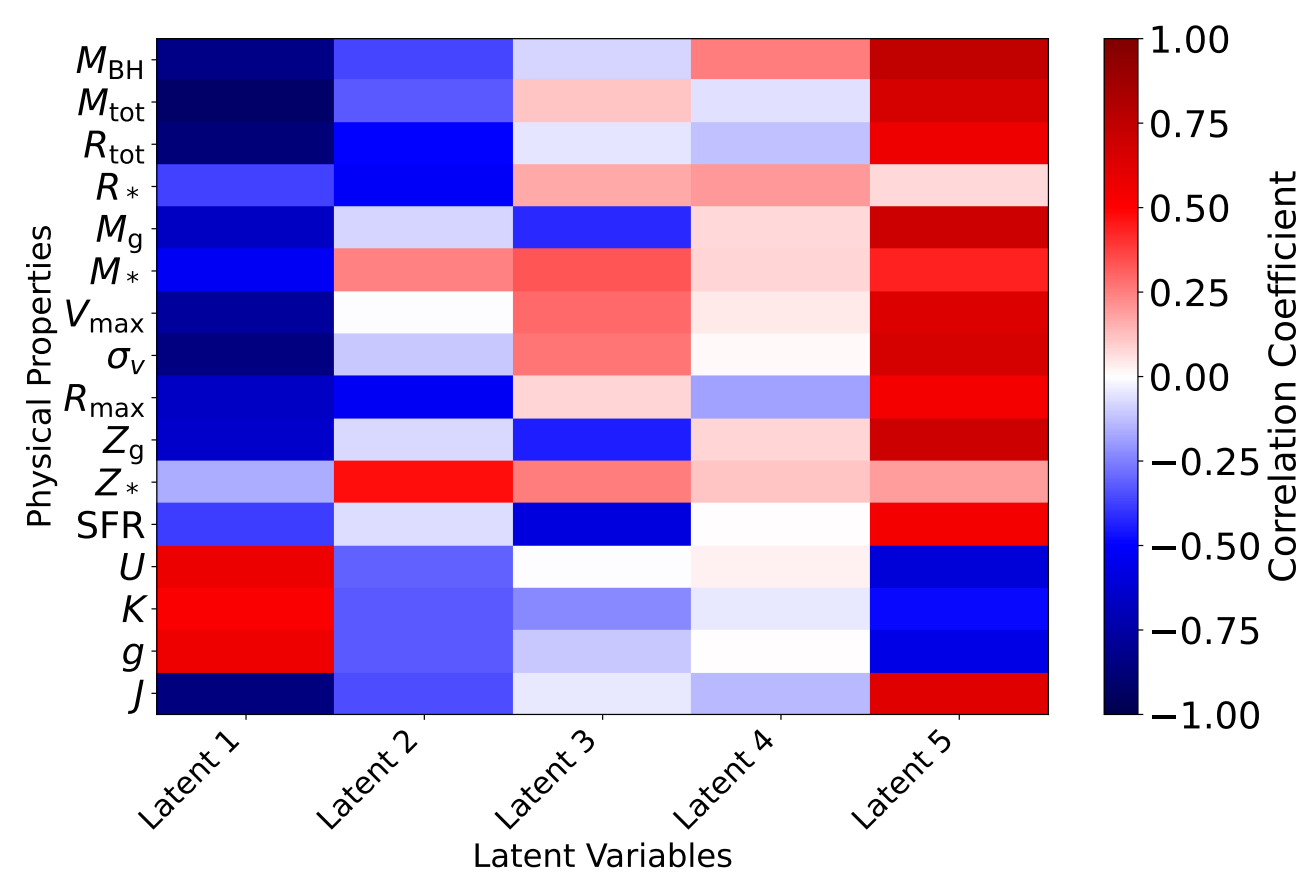

**Figure 5.** Linear correlation matrix between the top five latent variables and galaxy properties. The matrix shows correlations between the latent variables produced by training our model with an IOB layer at bottleneck width $k = 5$ and the input galaxy properties. The first and second latent variables exhibit strong linear correlations with specific input features, while the remaining latents show weaker or redundant patterns. As referenced in the main text, the "top latent variables" correspond to latents 1–5 shown in this figure.

In contrast, the lower latent variables demonstrate weaker linear correlations with galaxy properties or replicate trends already captured by earlier latent variables, such as the fifth latent variable mirroring the correlations of the first. This pattern suggests a hierarchical structure, where the top latent variables prioritize encoding the most essential information for reconstruction, while subsequent variables handle less critical or redundant details.

Additionally, we explored whether $\Omega_m$ is explicitly encoded in the latent variables. Given that our network achieves near-perfect reconstruction with $k = 16$, these 16 latent variables should, in theory, enable predictions of $\Omega_m$ with similar accuracy to the $\sim$10% precision reported by F. Villaescusa-Navarro et al. (2022), who utilized a fully connected neural network for this purpose. To test this, we first replicated their results using a smaller subset of 51 out of 1000 simulations. As shown in Figure 6 (left), even with this reduced training set, the network predicts $\Omega_m$ with approximately 10% precision.

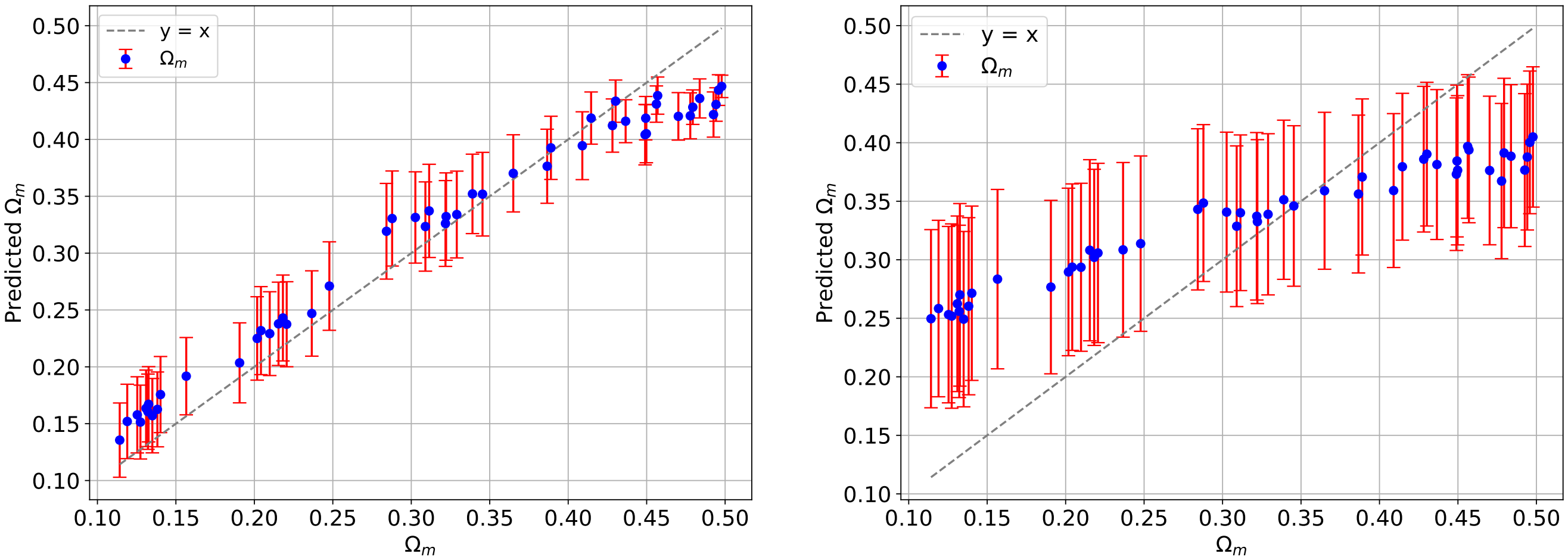

**Figure 6.** Comparison between the predicted and actual values of $\Omega_m$ obtained from the neural network proposed by F. Villaescusa-Navarro et al. (2022). The left panel displays results derived from a subset of CAMELS LH simulations, while the right panel showcases outcomes obtained through latent variables generated by an IOB autoencoder trained on CAMELS LH simulations. The increased error in the right panel suggests that $\Omega_m$ is not simply encoded in the latent variables, leading to a less precise prediction compared to the direct training approach.

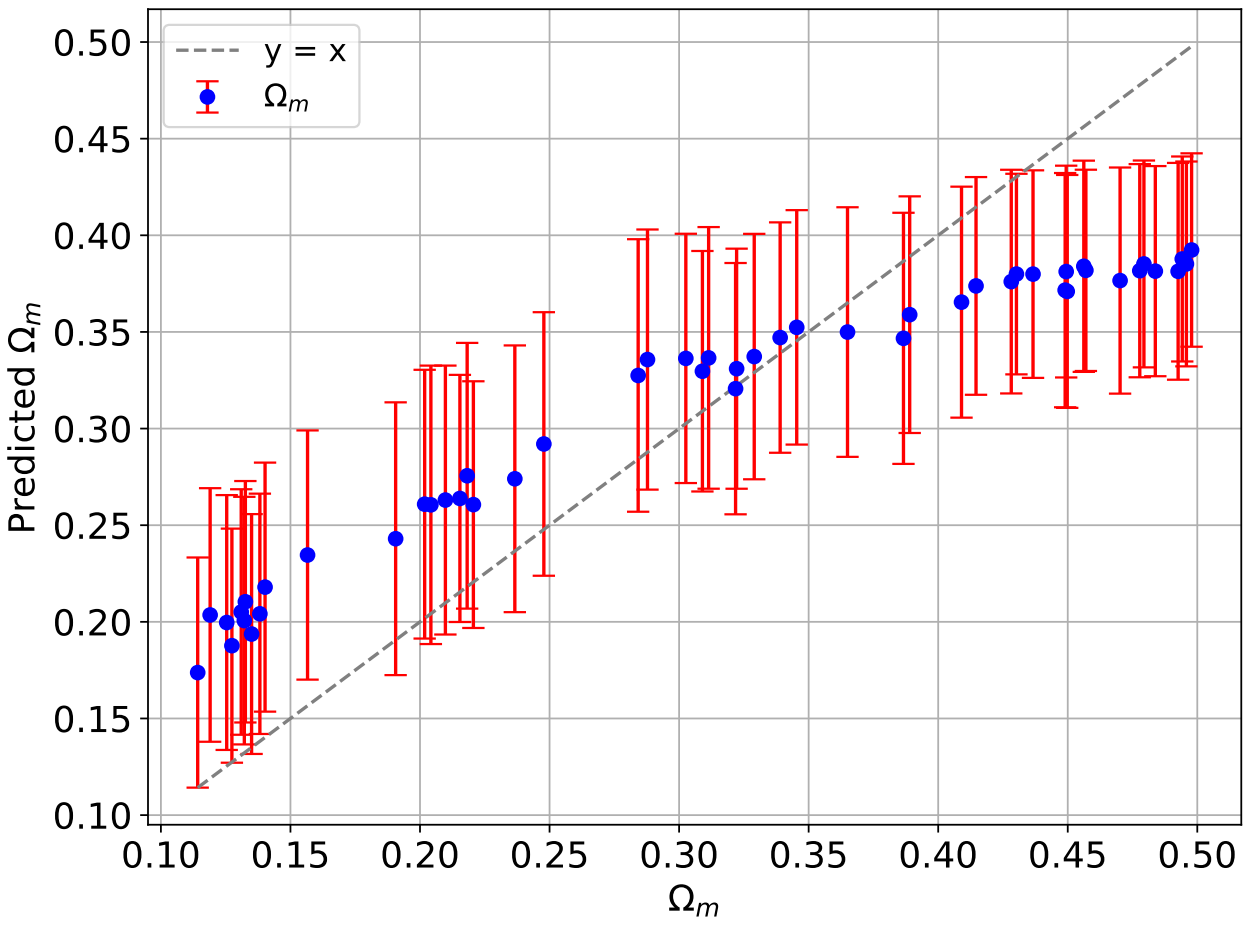

**Figure 7.** Comparison between the predicted and actual values of $\Omega_m$ obtained from the neural network proposed by F. Villaescusa-Navarro et al. (2022). The data had artificial noise added in covariance with the reconstruction errors of each galaxy property. The precision in predicting $\Omega_m$ significantly decreases when noise is present.

However, when training a network directly on the latent variables, we observe significantly higher errors, with the precision dropping to around 20%, as seen in Figure 6 (right). We hypothesize that this reduction in precision arises from the influence of the reconstruction errors, which act like noise. To investigate, we introduced small amounts of artificial Gaussian noise to the original data, calibrated to match the level of reconstruction error. This noise was structured according to the covariance in the reconstruction errors of individual galaxy properties. As shown in Figure 7, the addition of this noise further decreases the precision in predicting $\Omega_m$. This finding suggests that the signal for $\Omega_m$ may be masked by noise levels inherent in the latent representations.

At this stage, we cannot definitively determine the nature of the relationship between $\Omega_m$ and the manifold. It is possible that the manifolds corresponding to different $\Omega_m$ values are very close to each other, such that the added reconstruction errors scatter galaxies around the true manifold, obscuring the signal of $\Omega_m$. This scattering may account for the difficulty in predicting $\Omega_m$ directly from reconstructed galaxy properties.

## 5. Conclusion and Future Work

In summary, our study, building on the work of F. Villaescusa-Navarro et al. (2022), supports the existence of a low-dimensional manifold that captures the complex relationships within higher-dimensional galaxy properties space, extending beyond simple correlations. Using an autoencoder with an IOB layer, we explored how this manifold responds to variations in cosmological parameters, particularly $\Omega_m$, through the CAMELS 1P, LH, and CV simulation sets.

Our findings reveal that a model trained on the CV set, which has a fixed cosmology, struggles to reconstruct galaxies from the LH set. In the LH set, multiple parameters vary simultaneously, leading to increased reconstruction error with the fiducial model. When testing on galaxies in the 1P set, the correlation between parameter values and error is only evident when varying $\Omega_m$ or $A_{\mathrm{SN1}}$. This suggests these parameters significantly shift galaxy properties off the fiducial manifold learned by the model. In contrast, changes in $\sigma_8$, $A_{\mathrm{AGN1}}$, $A_{\mathrm{SN2}}$, and $A_{\mathrm{AGN2}}$ do not lead to significant deviations in reconstruction error, implying these parameters do not affect the manifold in the same way. This pattern persists even in the more complex LH simulations, reinforcing the conclusion that $\Omega_m$ and $A_{\mathrm{SN1}}$ are key drivers of shifts in galaxy properties.

Physically, the difference between $\Omega_m$ and $\sigma_8$ in influencing reconstruction error can be attributed to their distinct effects on galaxy formation (F. Villaescusa-Navarro et al. 2021). While $\sigma_8$ influences clustering and halo formation times, the CV set already spans a range of halo formation times, allowing the CV-trained model to learn to model variations in $\sigma_8$. In contrast, changes in $\Omega_m$ produce galaxies with properties fundamentally different from those in the CV set, resulting in higher reconstruction errors. Similarly, $A_{SN1}$, by influencing galactic wind energy, creates galaxies that deviate more from the fiducial set. However, $A_{SN2}$ primarily affects wind speed, which is influenced by local velocity dispersion and is already represented in the CV set, causing less deviation.

We also tested whether $\Omega_m$ could be directly encoded in the latent variables. Results from F. Villaescusa-Navarro et al. (2022) were reproduced with similar precision (10%), but when applying to latent variables from the IOB model, precision dropped to 20%, suggesting that noise may obscure the $\Omega_m$ signal in the latent space.

In future work, it would be interesting to investigate the role of noise further and its potential to introduce scatter or alter the manifold's dimensionality. This is particularly relevant when the reconstruction error is comparable to the manifold's extent, as it could cause galaxies to shift positions on the manifold, complicating predictions of $\Omega_m$. Applying these methods to observed galaxy properties, such as those from the MaNGA Galaxy Survey (N. Arora et al. 2023), would also be valuable. The MaNGA survey uses large-scale integral field spectroscopy to provide measurements of galaxy properties that previously could not be highly spatially resolved. In this case, an exploration of the role of noise would be of great importance as all observational data have some noise to be considered. Furthermore, this prompts an investigation into the idea of the robustness of the model as the manifold observed galaxy properties lie-on will be different from the manifold simulated galaxies lie-on.

## Acknowledgments

This work was supported by the Simons Collaboration on "Learning the Universe" and by NSF grant AST-2108678. The Flatiron Institute is supported by the Simons Foundation. We also acknowledge Lucy Reading-Ikkanda at the Simons Foundation for designing Figure 3, which visually captures the core concept of the galaxy property manifold.

## ORCID iDs

Amanda Lue https://orcid.org/0000-0002-4533-2581
Shy Genel https://orcid.org/0000-0002-3185-1540
Marc Huertas-Company https://orcid.org/0000-0002-1416-8483
Francisco Villaescusa-Navarro https://orcid.org/0000-0002-4816-0455
Matthew Ho https://orcid.org/0000-0003-3207-8868